\documentclass[entropy,article,accept,moreauthors,pdftex,12pt,a4paper]{mdpi}

\lastpage{556}
\doinum{10.3390/e16010543}
\pubvolume{16}
\pubyear{2014}
\history{Received: 01 December 2013; in revised form: 16 December 2013   / Accepted: 30 December 2013 / \\
Published: 16 January 2014}
\pdfoutput=1

\usepackage{graphicx}
\usepackage{epstopdf}
\usepackage{soul}
\Title{Entropy and the Predictability of Online Life}

\Author{Roberta Sinatra $^{1}$ and Michael Szell $^{2,}$*}

\address{%
$^{1}$ Center for Complex Networks Research and Physics Department, Northeastern University, \mbox{110 Forsyth Street}, Boston, MA 02115, USA; E-Mail: r.sinatra@neu.edu\\
$^{2}$ Senseable City Laboratory, Massachusetts Institute of Technology, 77 Massachusetts Avenue, Cambridge, MA 02139, USA}

\corres{E-Mail: mszell@mit.edu; \linebreak Tel.: +1-617-324-4474; Fax: +1-617-258-8081.}

\abstract{Using mobile phone records and information theory measures, our daily lives have been recently shown to follow strict statistical regularities, and our movement patterns are, to a large extent, predictable. Here, we apply entropy and predictability measures to two datasets of the behavioral actions and the mobility of a large number of players in the virtual universe of a massive multiplayer online game. We find that movements in virtual human lives follow the same high levels of predictability as offline mobility, where future movements can, to some extent, be predicted well if the temporal correlations of visited places are accounted for. Time series of behavioral actions show similar high levels of predictability, even when temporal correlations are neglected. Entropy conditional on specific behavioral actions reveals that in terms of predictability, negative behavior has a wider variety than positive actions. The actions that contain the information to best predict an individual's subsequent action are negative, such as attacks or enemy markings, while the positive actions of friendship marking, trade and communication contain the least amount of predictive information. These observations show that predicting behavioral actions requires less information than predicting the mobility patterns of humans for which the additional knowledge of past visited locations is crucial and that the type and sign of a social relation has an essential impact on the ability to determine future behavior.}


\begin{document}

%%%%%%%%%%%%%%%%%%%%%%%%%%%%%%%%%%%%%%%%%%

\vspace{-12pt}
\section{Introduction}

Capturing the regularities of our daily lives and the occasional deviations from the steady diurnal patterns has traditionally eluded an all-encompassing approach, due to tremendous efforts in monitoring detailed human activities over long times and the bias in behavior caused by obtrusive methods of observation \cite{rosenthal1991mps}. However, the recent ability to address questions in social science by using huge datasets that have emerged over the past decades as a result of digitalization has opened previously unimaginable ways of conducting research in the field \cite{lazer2009css}.

On the one hand, these new datasets give a highly detailed protocol of our ordinary lives, for example, in the form of mobile phone data, which enables a deeper understanding of the regularities in our mobility patterns \cite{gonzalez2008uih,schneider2013udh}, and how the regularities in human behavior are reflected in the geographic regions that emerge from our interactions \cite{sobolevsky2013dgr}. On the other hand, from a previous point of view, ``extraordinary'' new forms of human behavior can now be observed online, where the full set of all actions performed in the system is typically available for study, spanning an even deeper level of detail. Online social networking services, such as Twitter or Facebook, or discussion forums allow new insights into the rhythms of social actions and interactions, as expressed online \cite{golder2011dsm,golder2007rsi,mitrovic2010bloggers,tadic2013cme}, and how these interactions relate to the underlying offline events \cite{szell2013cor}. An even richer insight can be gained into human-led lives that unfold \emph{entirely} in artificial online environments, such as in persistent, massive multiplayer online games, where human-controlled characters spend their whole virtual lives within an online world interacting with other characters \cite{bainbridge2007srp}. The playing of online games is one of the most wide-spread forms of collective human behavior in the world; the ``massive multiplayer'' aspect allows one to not only study single individuals, but also collective behavioral phenomena that typically emerge in complex social systems \cite{ball2004pmh}. Here, data can be available of all actions, decisions and interactions between many thousands of individuals over long time spans \cite{szell2012sdl}, allowing understanding of the structure and evolution of socio-economic \mbox{networks \cite{szell2010mol,szell2010msd,klimek2013tcd}}, mobility \cite{szell2012ums} or the emergence of good conduct \cite{thurner2012egc} and elite structures \cite{corominas2013des} in large social systems.

Going hand in hand with the new availability of large-scale, longitudinal behavioral datasets of various kinds, well-known methods from the mathematical and physical sciences, especially statistical physics and information theory \cite{castellano2009sps,sinatra2010nms,gallotti2012tsp}, have been extended and/or re-applied successfully in this context. In particular, principal component analysis and the concept of ``eigenbehavior'' has been used to quantify behavioral regularities and to predict future activities in the daily lives of a group of 100 subjects \cite{eagle2009eis}. Similarly, information theory measures provide an adequate quantification between uniform distributions (maximal entropy) and maximally uneven distributions of states (minimal entropy), which, in the case of human behavior, can inform us about the extent of uniformity and, thus, predictability in our activity patterns. The concept of entropy has been applied specifically to assess the predictability of mobility patterns \cite{song2010lph,gallotti2013emi}, of economic behavior \cite{krumme2010pis}, the order of human-built structures, such as urban street networks \cite{gudmundsson2013eou}, or the complexity of online chatting behavior \cite{takaguchi2011pcp,wang2012hro,tadic2013cme}. Further, a theoretical framework for non-extensive entropies has been recently developed that might be well applicable to\linebreak complex systems \cite{hanel2011ccc,hanel2012generalized}.

\newpage
%%%%%%%%%%%%%%%%%%%%%%%%%%%%%%%%%%%%%%%%%%

\section{Behavior and Mobility Data of Human Players in the Online World, Pardus}

Here, our goal is to apply classical entropy measures to study the patterns of various kinds of behavior in a single, closed socio-economic system, as generated by thousands of users in the online game ``Pardus'' \cite{szell2012sdl}, to provide an insight into the regularity of life in online worlds and, eventually, to draw possible conclusions on how humans lead their offline lives.

\subsection{The Online World, Pardus}

The online world Pardus, www.pardus.at, is a browser-based, massive multiplayer online game open to the public for over nine years. Over 400,000 users have registered to play so far. The game features three independent, persistent game universes, which had a defined starting time, but no scheduled end. There are no predefined goals in the game: many aspects of social life within Pardus are self-organized, for example, the emergence of social groups (alliances) and the politics between them. Players are engaged in a multitude of social activities, \textit{i.e.},~chatting, cultivating friendships, building up alliances, but also negative interactions, such as destructive attacks, and economic activities, such as producing commodities in factories and selling them to other players. We focus on the ``Artemis'' game universe, in which we recorded player actions over the first 1,238 consecutive days of the universe's existence. Communication between any two players can take place directly, by using a one-to-one, e-mail-like private messaging system. We focus on one-to-one interactions between players only and discard indirect interactions, such as, e.g., participation in chats or forums. There are global interactions, \textit{i.e.},~interactions that can be performed independently of the spatial position of players in the game universe, which are communication, setting and removing friendship or enemy links, or placing a bounty on another player. The actions of trade or attack, however, need players to meet in space. All data used in this study are \mbox{fully anonymized.}

\subsection{Mobility Time Series}

For studying regularities in mobility patterns, we use the same dataset of Pardus player movements that has been used in \cite{szell2012ums}. A universe in Pardus can be represented as a network with 400 nodes, called sectors, and 1,160 links. Each sector is like a city, where players can have social relations or entertain economic activities. Typically, sectors adjacent on the universe map, as well as a few far-apart sectors, are interconnected by links that allow players to move from sector to sector. At any point in time, each sector is usually attended by a large number of players. The universe network has a large diameter of 27, which means that, on average, players have to move through a non-negligible number of sectors to traverse the universe. Due to a limited pool of actions that players can spend on movement, traveling large distances can take a player several days. Using this dataset, we previously studied the statistical movement patterns of players and found that locations are visited in a specific order, leading to strong long-term memory effects \cite{szell2012ums}. In detail, we extract player mobility data from day 200 to day 1,200 of the universe's existence. We discard the first 200 days, because social networks between players of Pardus have shown aging effects in the beginning of the universe \cite{szell2010msd}. To make sure we only consider active players, we select all who exist in the game between the days 200 and 1,200, yielding 1,458 players active over a time-period of 1,000 days. The sector IDs of these players, \textit{i.e.}, their positions on the universe network's nodes, are logged every day at 05:35 GMT%please define
.

\subsection{Behavioral Action Time Series}

For studying regularities in behavioral time series, we use the same dataset of Pardus player actions that has been used in \cite{thurner2012egc}. Players can express their sympathy (distrust) toward other players by establishing so-called friendship (enmity) links. These links are only seen by the player marking another as a friend (enemy) and the respective recipient of that link. For more details on the game, see \cite{szell2010msd}. \linebreak We consider eight different actions every player can execute at any time. These are communication (C),\linebreak trade (T), setting a friendship link (F), removing an enemy link (forgiving) (X), attack (A), placing a bounty on another player (punishment) (B), removing a friendship link (D) and setting an enemy \mbox{link (E).} While C, T, F and X can be associated with positive actions, A, B, D and E are hostile or negative actions. We classify communication as positive, because only a negligible part of communication takes place between enemies \cite{szell2010msd}. Following a previous formalism \cite{szell2010msd}, we say that positive actions have a positive sign, and negative actions have a negative sign. The alphabet, $\mathcal{X}$, of all possible dyadic actions happening in each player's life therefore spans 16 letters: eight possible performed actions (four negative, \mbox{four positive}) and eight possible received actions (four negative, four positive). We denote received actions with the suffix, $r$, e.g.,~$A_r$ for a received attack. Due to the heterogeneous activity patterns of players, we operate in action-time rather than in actual time; for example, indices of $t$ and $t-1$ denote that two actions were subsequent, regardless of whether the actual time difference was seconds or weeks \cite{thurner2012egc}. From all sequences of all \mbox{34,055 Artemis} players who performed or received an action at least once within 1,238 days, we removed players with a life history of less than 1,000 actions, leading to the set of the most active 1,758 players that are considered throughout this work.

%%%%%%%%%%%%%%%%%%%%%%%%%%%%%%%%%%%%%%%%%%

\section{Entropy and Predictability Measures}

To study the regularity and predictability of behavior from the discrete time series, we use three entropy measures. Following \cite{song2010lph}, we call the binary logarithm of the number of distinct states, $N_i$, of a player, $i$, the \emph{random entropy}:
\begin{equation}
S^{\mathrm{rand}}_{i} = \log_{2} N_{i}\label{eq:srand}
\end{equation}
In the case of mobility, ``states'' refer to the 400 possible sectors in the universe visitable at a given point in time by a player. The maximal possible random entropy is $S^{\mathrm{rand}} = \log_{2} 400 \approx 8.6$, reached when all sectors are visited at least once. In the case of behavioral actions, a state can be one of the 16 possible action or received action types; here, the maximum possible random entropy is $S^{\mathrm{rand}} = \log_{2} 16 = 4$.

The Shannon entropy, $S^{\mathrm{unc}}_{i}$, of a player, $i$, is defined as:
\begin{equation}
S^{\mathrm{unc}}_{i} = - \sum_{x \in \mathcal{X}_i}p_{i}(x)\log_{2} p_{i}(x)\label{eq:sunc}
\end{equation}
where $p_{i}(x)$ is the measured probability over the respective time span that player $i$ has occupied a state, $x$, and $\mathcal{X}_i$ is the ensemble of the $N_i$ distinct states. In this context, we call the Shannon entropy the \mbox{\emph{temporal-uncorrelated entropy}}, because it captures the entropy when the temporal order of states is ignored \cite{song2010lph}. The random and temporal-uncorrelated entropies are equal, $S^{\mathrm{rand}}_{i} = S^{\mathrm{unc}}_{i}$, if all of the $N_i$ distinct states, $x$, were occupied with uniform probability $p_{i}(x) = 1/N_i$ by the player, $i$. For mobility, the occupation of a single sector over the whole time span of 1,000 days would result in the smallest possible random and temporal-uncorrelated entropy of $S^{\mathrm{rand}} = S^{\mathrm{unc}} = 0$.

Finally, we make use of the conditional entropy $S^{\mathrm{cond}}_{i}$ of a player, $i$, capturing the entropy conditional on temporal short-term correlations over one previous state in the time series,
\begin{equation}
S^{\mathrm{cond}}_{i} = - \sum_{x_t \in \mathcal{X}_i}\sum_{x_{t-1} \in \mathcal{X}_i}p_{i}(x_{t-1},x_t) \log_{2} p_{i}(x_t|x_{t-1})\label{eq:scond}
\end{equation}
with $p_{i}(x_{t-1},x_t)$ being the probability of occurrence of the pair of subsequent states, $x_{t-1}$ and $x_t$, $p_{i}(x_t|x_{t-1}) = p_{i}(x_{t-1},x_t) / p(x_{t-1})$, the probability of the state, $x_t$, at time $t$ given a preceding state, $x_{t-1}$. The conditional and temporal-uncorrelated entropies are equal, $S^{\mathrm{cond}}_{i} = S^{\mathrm{unc}}_{i}$, if there are no \mbox{temporal correlations.} %In case of a Markov chain, i.e.,~in the absence of long-term temporal correlations, the conditional entropy would coincide with the actual entropy.

It is easy to show that we have $S^{\mathrm{cond}} \leq S^{\mathrm{unc}} \leq S^{\mathrm{rand}}$ for each user \cite{Cover:2006vu}. The differences in these two inequalities quantify the effects of short-term temporal correlations and the uniformity of the occupation distribution, respectively. To assess the predictability of specific states or of classes of states, we also define the conditional entropy for the set of states, $\mathcal{Z}$,
\begin{equation}
S^{\mathrm{cond}}_{i}(\mathcal{Z}) = - \sum_{x_t \in \mathcal{X}_i}\sum_{x_{t-1} \in \mathcal{X}_i \cap \mathcal{Z}}p_{i}(x_{t-1},x_t) \log_{2} p_{i}(x_t|x_{t-1})\label{eq:scond2}
\end{equation}
which is the conditional entropy given that the previous state belonged to $\mathcal{Z}$, where $\mathcal{Z}$ can be fixed as any subset of all the possible states, $\mathcal{X}$. Notice that $S^{\mathrm{cond}}_{i} \equiv S^{\mathrm{cond}}_{i}(\mathcal{Z}) + S^{\mathrm{cond}}_{i}(\mathcal{X}_i \setminus \mathcal{Z})$.

Complementary to entropy measures of information content or \emph{unpredictability} are measures of \emph{predictability} that denote in a percent value how likely an appropriate predictive algorithm could foresee an individual's future behavior \cite{song2010lph}. The predictability, $\Pi_i^{\bullet}$, of an individual $i$ is bounded above by:
\begin{equation}
S_i^{\bullet} = H(\Pi_i^{\bullet}) + (1-\Pi_i^{\bullet}) \log_2 (N_i-1)\label{eq:predictability}
\end{equation}
with the binary entropy function:
\begin{equation}
H(\Pi_i^{\bullet}) = \Pi_i^{\bullet} \log_2(\Pi_i^{\bullet}) - (1-\Pi_i^{\bullet}) \log_2(1-\Pi_i^{\bullet})
\end{equation}
where $\bullet$ is a placeholder for any of the types, $\mathrm{rand}$, $\mathrm{unc}$ or $\mathrm{cond}$. Unlike the measure of entropy, which is well established, the application of this predictability measure to practical problems is relatively recent. It is based on the idea that predictability is related to the error probability in guessing the outcome of a discrete random variable \cite{feder1994ree}. The upper bound given in Equation~(\ref{eq:predictability}) comes from Fano's \mbox{inequality \cite{fano1961ti,Cover:2006vu}.}  For a detailed discussion on this bound and on possible lower bounds, see \cite{feder1994ree,song2010lph}.

For being able to study in more detail the effects of memory in the system \cite{sinatra2011maximal,chierichetti2012awu}, we generalize the conditional entropy:
\begin{equation}
S^{\mathrm{cond},k}_{i} = - \sum_{x_t \in \mathcal{X}_i}\cdots \!\sum_{x_{t-k} \in \mathcal{X}_i} p_{i}(x_{t-k},\ldots,x_t)\log_{2} p_{i}(x_t|x_{t-k},\ldots,x_{t-1})\label{eq:sunck}
\end{equation}
where $k$ is an integer denoting the memory window. Note that $S^{\mathrm{\mathrm{cond},1}}_{i} \equiv S^{\mathrm{cond}}_{i}$ and that we can identify $S^{\mathrm{\mathrm{cond},0}}_{i}$ with $S^{\mathrm{unc}}_{i}$. It follows from Fano's inequality \cite{fano1961ti} that $S^{\mathrm{\mathrm{cond},1}}_{i} \geq S^{\mathrm{\mathrm{cond},2}}_{i} \geq S^{\mathrm{\mathrm{cond},3}}_{i} \geq \cdots$. \linebreak The differences between subsequent values in this chain inform us about the gain of predictability when we increase the memory window one by one. If such a difference starts becoming negligible from a particular level, $k$ to $k+1$, it means that the system does not exhibit relevant memory effects beyond a window of $k$ steps. If this level is at $k=0$, the events are uncorrelated; if at $k=1$, the system is Markovian, otherwise, it is non-Markovian.

%%%%%%%%%%%%%%%%%%%%%%%%%%%%%%%%%%%%%%%%%%

\section{Results and Discussion}
\vspace{-12pt}

\subsection{Predictability in Mobility}

\begin{figure}[H]
\centering
 \includegraphics{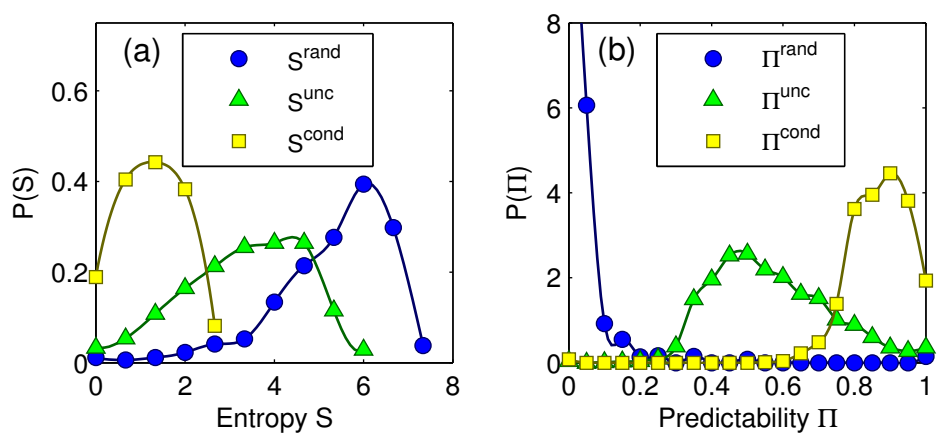}
 \caption{The distribution of ({\bf a}) entropy and ({\bf b}) the predictability measures of the mobility of the Pardus players. Both are almost identical to the mobility of humans in geographic space \cite{song2010lph}: Each considered entropy measure improves predictability substantially, from considering the uniformity of occupation to additionally short-term temporal correlations.\label{fig:entropymobility01}}
\vspace{-12pt}
\end{figure}

We applied all entropy and predictability measures to the mobility time series, Figure~\ref{fig:entropymobility01}a,b, respectively. Results show almost identical predictability behavior for humans in our online world as for the mobility of humans in geographic space \cite{song2010lph,gallotti2012tsp}. The distributions for $S^{\mathrm{unc}}$ and $S^{\mathrm{rand}}$ are both qualitatively and quantitatively matching, showing that also online, movements of human avatars have the same highly predictable patterns when temporal correlations are accounted for, but are mostly unpredictable when the order of visitations is ignored. In particular, also here, $S^{\mathrm{rand}}$ peaks around six, indicating that an individual who chooses her next location randomly could be found, on average, in any of $2^{S^{\mathrm{rand}}} \approx 64$ locations, which is a substantial part of the 400 possible sectors. The contrasting peak of $S^{\mathrm{cond}}$ below two shows that the actual uncertainty of a typical player's location is not 64, but rather, less than $2^2 = 4$ sectors. The conditional entropy, $S^{\mathrm{cond}}$, is not directly comparable to the actual entropy, $S$, in \cite{song2010lph}, but shows the same tendency in that temporal correlations are substantial, even if just having a memory of one. However, for re-creating the statistical features of mobility thoroughly, longer memory is needed \cite{szell2012ums}. The peak of $\Pi^{\mathrm{cond}}$ around $0.9$ means that only in around $10\%$ of cases does a player choose her location in a manner that appears to be random, but in $90\%$ of the cases, we can hope to predict her whereabouts with an appropriate predictive algorithm. This high predictability stands in contrast to the moderately predictive case given by $\Pi^{\mathrm{unc}}$ peaking around $0.5$ and the highly unpredictive case of $\Pi^{\mathrm{rand}}$ peaking narrowly and close to zero.

\subsection{Predictability in Behavioral Actions}

\begin{figure}[H]
\centering
 \includegraphics[scale=.9]{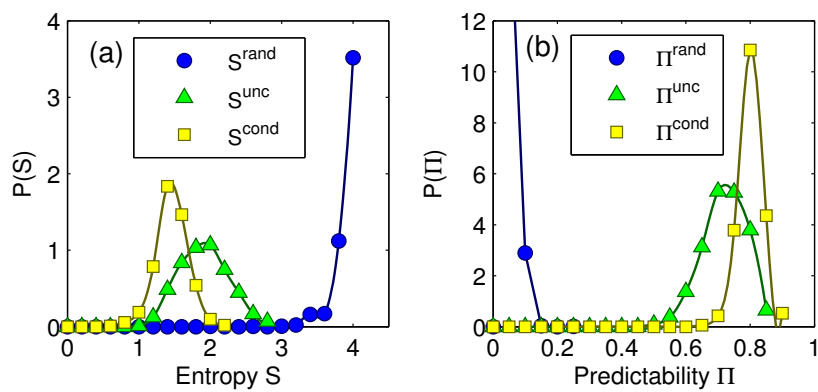}
 \caption{Distribution of ({\bf a}) entropy and ({\bf b}) predictability measures of the behavioral actions of the Pardus players. As in the case of mobility, behavioral actions are highly regular and predictable. However, the predictability gained from considering the uniformity of occupation is much larger than the predictability gained from also considering \mbox{temporal correlations}.\label{fig:entropybehavior01}}
\end{figure}

A similar picture to mobility arises for behavioral actions. Figure~\ref{fig:entropybehavior01}a,b, respectively, report the entropy and predictability distributions of all 16 types of actions and received actions. Here, $S^{\mathrm{rand}}$ is peaked at four, showing that most players are making full use of their behavioral possibilities of $16 = 2^4$ action and received action types in the course of their online lives. However, the sharp drop to the distribution of $S^{\mathrm{unc}}$, which peaks around two, shows that, in practice, most of these actions and received actions are focused on around $2^2 = 4$ action or received action types only. The even narrower curve of $S^{\mathrm{cond}}$, which peaks around $1.5$, with a corresponding peak of $\Pi^{\mathrm{cond}}$ at $0.8$, demonstrates that the conditional information allows us to predict $80\%$ of actions. This is only slightly more than the $73\%$ prediction rate peak from $\Pi^{\mathrm{unc}}$; however, $\Pi^{\mathrm{cond}}$ is distributed more widely. In conclusion, the predictability gained from considering the uniformity of occupation is much larger than the predictability gained from also considering Markovian temporal correlations, as opposed to the case of mobility where temporal correlations add substantial predictive value.

\begin{figure}[H]
\centering
 \includegraphics[scale=.9]{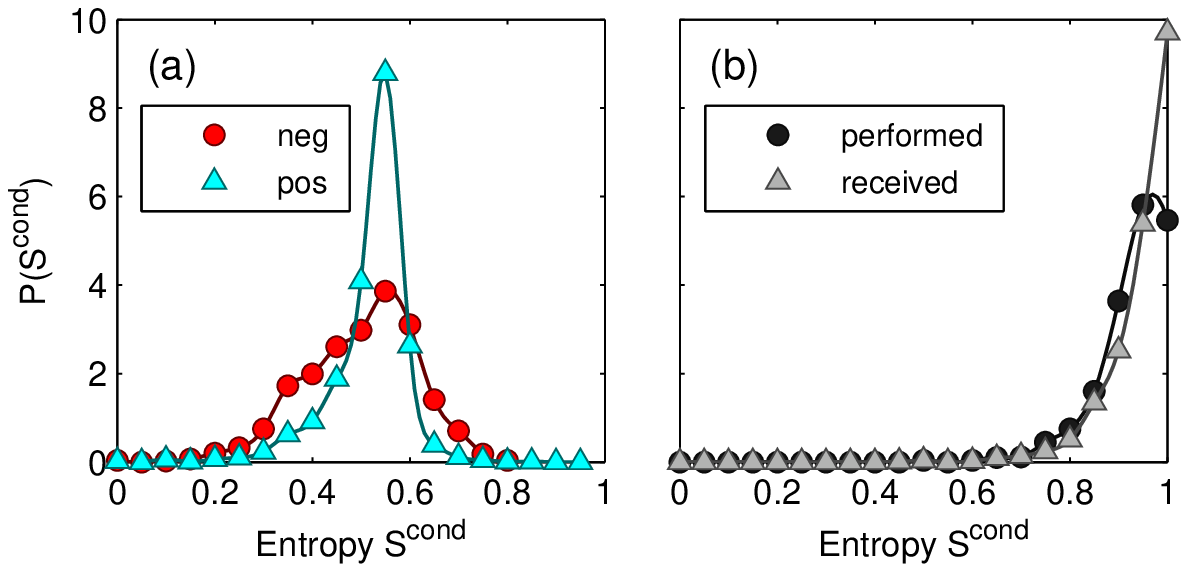}
 \caption{The distribution of the conditional entropy measures of the behavioral actions of the Pardus players, given that the previous action belonged to a certain category. ({\bf a}) Entropy given that the previous action or received action was positive/negative. The positive and negative distributions have their maxima both around $0.55$, but the former is much more narrow than the latter one, showing that there is a much wider range of negative behavior in terms of predictability than positive behavior. ({\bf b}) Entropy given that the previous action was performed/received. Both distributions peak very close to one, showing that the information of whether an action was performed or received does, in general, not have a high predictive value. \mbox{The peak for} the received actions is slightly closer to one than for the performed actions. \label{fig:entropyposneg01}}
\end{figure}

One previous key observation on Pardus players is the fundamental structural and dynamic difference between positive and negative action types and their interaction networks \cite{szell2010mol,szell2012sdl,szell2010msd,thurner2012egc}. To see if this difference is also apparent in the extent of predictability, we plotted the distribution of the conditional entropy of the players given that the previous action or received action was positive/negative (Figure~\ref{fig:entropyposneg01}a), \textit{i.e.},~the set $\mathcal{Z}$ in Equation~(\ref{eq:scond2}) corresponds to $\mathcal{Z}=\{C,T,F,X,C_r,T_r,F_r,X_r\}$ or to $\mathcal{Z}=\{A,B,D,E,A_r,B_r,D_r,E_r\}$, respectively. We aim to understand whether the actions that follow positive actions are more predictable than those that follow negative actions. If the distributions were identical, the sign of an action would cause no difference in the predictability of the subsequent action. In fact, although both distributions peak around $0.55$, showing that there is a moderate amount of predictive value gained from the information of an action's sign, the positive distribution is much more narrow than the negative one, implicating that there is a much wider range of negative behavior in terms of predictability than positive behavior. This result suggests that ``good'' people are much alike, but ``bad'' persons behave badly in more various and, sometimes, more unpredictable ways.

The conditional entropy for performed or received actions, \textit{i.e.},~$\mathcal{Z}=\{C,T,F,X,A,B,D,E\}$ or $\mathcal{Z}=\{C_r,T_r,F_r,X_r,A_r,B_r,D_r,E_r\}$ in Equation~(\ref{eq:scond2}), respectively, is peaked very narrowly and close to one for both cases and slightly more so for received actions; Figure~\ref{fig:entropyposneg01}b. This observation shows that the directionality of actions contains much less predictive information than the sign of an action.

\begin{figure}[H]
\centering
 \includegraphics[scale=.75]{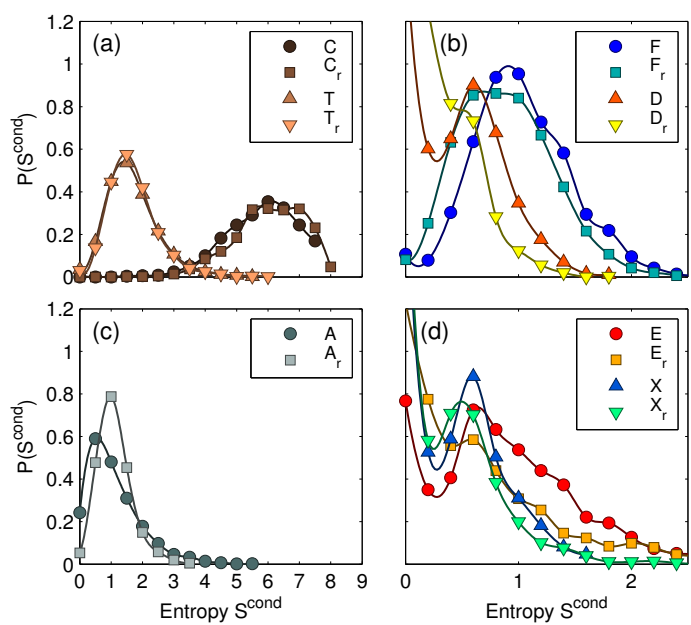}
 \caption{The distribution of conditional entropy measures of the behavioral actions of the Pardus players, given that the previous action was of a certain type. ({\bf a}) The distributions for performed and received communication events ($C$ and $C_r$) and for performed and received trade events ($T$ and $T_r$). Communication peaks around six bits, trade around $1.3$ bits. Performed and received actions do not show substantial deviations here. ({\bf b}) The distributions \protect\linebreak for performed and received friendship marking events ($F$ and $F_r$) and for performed and received friendship removals ($D$ and $D_r$). The curves peak around one or lower. \protect\linebreak {({\bf c}) The distributions} for performed and received attacks ($A$ and $A_r$). The former curve peaks below one; the latter peaks around one and is narrower. ({\bf d}) The distributions for performed and received enemy marking events ($E$ and $E_r$) and for performed and received enemy removals ($X$ and $X_r$). All the curves peak once around $0.6$ and another time close to zero.\label{fig:entropyspecific01}}
\vspace{-16pt}
\end{figure}

We can further refine the conditional entropy measure by considering single actions as the condition, \textit{i.e.},~where $\mathcal{Z}$ in Equation~(\ref{eq:scond2}) is a singleton, to assess how much each action or received action type allows one to predict the subsequent action that is about to happen in a player's life. The conditional entropy of trade peaks around $1.3$; the distribution of communication is more wide, peaking around six bits; Figure~\ref{fig:entropyspecific01}a. Distributions of received trades and communications are almost identical, only received communication is slightly more right-skewed than performed actions of communication. The reason why communication is associated with higher unpredictability might have to do with the game's action point system \cite{szell2010msd}: every action, except the action of communication, costs an amount of so-called action points for which every player has only a limited pool. Therefore, players are not limited in their communication behavior, but are so for trade, friendship markings, \textit{etc}. The entropy distribution of friendship marking, $F$, Figure~\ref{fig:entropyspecific01}b, peaks around one bit and is, therefore, much less unpredictable. The entropy of enemy marking $E$ peaks even closer to zero (Figure~\ref{fig:entropyspecific01}d); all of the actions related to enemy markings, $E$, $E_r$, $X$ and $X_r$, show a bimodal distribution with an extra peak at zero, but this is clearly not the case for friendship markings $F$ or $F_r$. This bimodality could hint towards two different kinds of effects that arise from enemy marking, where, for example, either the person who makes or removes the marking immediately predictably sends a message to the recipient in a fraction of cases, or in the remaining fraction, this does not happen. Finally, the conditional entropy of received attacks, $A_r$, peaks around one, and performed attacks, $A$, are more wide peaking at a smaller value; Figure~\ref{fig:entropyspecific01}c. In all the distributions that deal with friendship or enemy markings, $F$, $D$, $E$ and $X$, we observe a right-shift of peaks for received actions, meaning that a player's next action is more predictable given that a friend/enemy event happened to her, as opposed to when she performed such an action towards somebody else. For attacks, however, we see the opposite. It is unclear what causes this phenomenon or how relevant it is: we can only speculate that a received attack could have a possibly stronger emotional impact on a player and, therefore, a more adverse effect on the predictability on her next action, while this is \textit{vice versa} for friendship/markings. Further, it is interesting to note that the removal of a friendship link has a similar pattern to the addition of an enmity link, suggesting that these two actions might be closely related, since they have a similar impact on future behavior. In general, however, the removal of a positive/negative tie cannot always be put on the same level as the addition of a negative/positive tie, as the reversed case of friendship addition and enemy removal shows.

\begin{figure}[H]
\centering
 \includegraphics[scale=1.2]{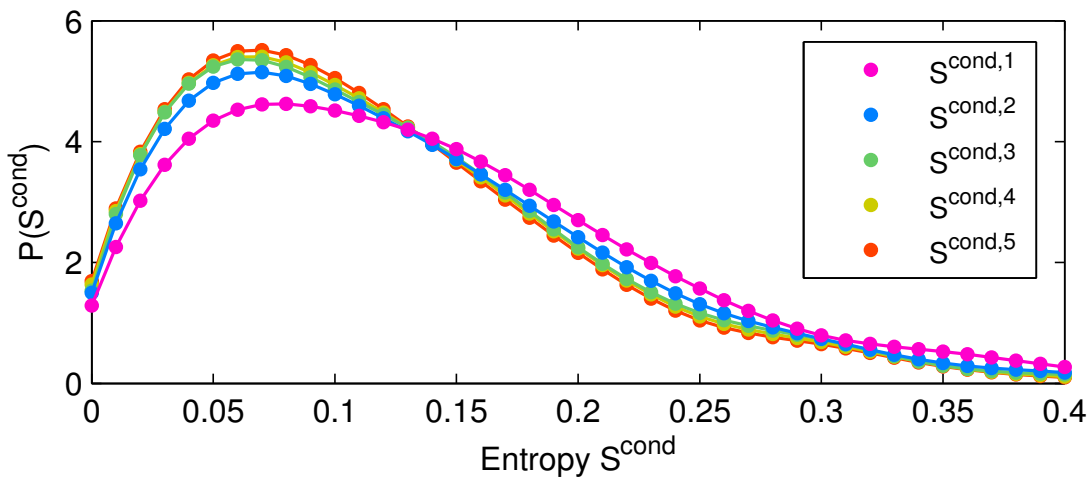}
 \caption{The convergence of the conditional entropy of the positive and negative behavioral actions of Pardus players with an increasing memory window. The difference between $S^{\mathrm{cond},1}$ and $S^{\mathrm{cond},2}$ is small, $D(2) = 0.0097$, showing that the system is almost Markovian. \protect\linebreak For higher memory windows, we have $D(3) = 0.0020$, $D(4) = 0.0006$ and $D(5) = 0.0005$, indicating almost identical distributions, which implies that there are practically no long-term correlations in the signs of behavioral actions.\label{fig:entropycondk}}
\end{figure}

Finally, we are interested in assessing the memory dependence of the behavioral actions in the \linebreak system \cite{chierichetti2012awu},  \textit{i.e.},~the gain of predictability from conditional entropies with longer time windows, using the measures, $S^{\mathrm{cond},k}$, for increasing $k$. Unfortunately, in practice, these rely on the empirical \linebreak probabilities, $p_{i}(x_{t-k},\ldots,x_t)$, of all possible substrings $x_{t-k},\ldots,x_t$ (see Equation~(\ref{eq:sunck})), which would lead to combinatorial explosion with our alphabet size of 16. For example, $k=3$ would mean $16^3 = 4,096$ possible substrings of a length of three, many of which do not exist at all or are statistically not reliable to assess from a dataset of 1,758 players, each having performed up to a few thousand actions. Therefore, in the following, we used the simplified alphabet of a size of two of negative or positive actions, allowing feasible calculation of $S^{\mathrm{cond},k}$ up to $k=5$. The distributions of these entropies are shown in Figure~\ref{fig:entropycondk}. The distributions converge quickly, showing only a small difference between $S^{\mathrm{cond},1}$ and $S^{\mathrm{cond},2}$ and almost no difference between higher order distributions. We quantify these differences via the Kullback--Leibler divergence between the distributions of the conditional entropy of subsequent memory levels, $S^{\mathrm{cond},k-1}$ \linebreak and $S^{\mathrm{cond},k}$,
\begin{equation}
D(k) = D(S^{\mathrm{cond},k} || S^{\mathrm{cond},k-1}) = \sum_{j} S^{\mathrm{cond},k}(j) \log \frac{S^{\mathrm{cond},k}(j)}{S^{\mathrm{cond},k-1}(j)}\label{eq:kullbackleibler}
\end{equation}
which provides the information gain for going from a memory of a length of $k-1$ to $k$ \cite{Cover:2006vu,sinatra2011maximal}. A divergence of zero means that two distributions are identical. The first values from $D(2)$ to $D(5)$ read $0.0097$, $0.0020$, $0.0006$ and $0.0005$. For comparison, the Kullback--Leibler divergence between $S^{\mathrm{unc}}$ and $S^{\mathrm{cond}}$, $D(1)$, yields the much higher value of $0.38$, showing that the system is, to a large part, Markovian and that the predictability gained from higher-order correlations is negligible.

%%%%%%%%%%%%%%%%%%%%%%%%%%%%%%%%%%%%%%%%%%

\section{Conclusions}

We applied three measures of entropy to two sets of time series of the behavioral actions and the movements of a large number of players in a virtual universe of a massive multiplayer online game. \mbox{We found that} movements in virtual human lives follow identical levels of predictability as offline mobility. This result reasserts previous observations on the similarities between the online and offline movements of humans \cite{szell2012ums} and is especially striking considering that in online worlds, individuals are not performing physical movements, but rather, navigate a virtual avatar.

Extending the approach to behavioral time series, also, here, we were able to provide evidence for high predictability. However, in this case, we found that due to weaker temporal correlations, there is hope to more easily predict behavioral actions than the temporally correlated mobility patterns of humans for which information about previously visited locations is required. Findings using entropy measures conditional on positive and negative actions suggest that ``good'' people are much alike, but ``bad'' persons behave badly in more various and, sometimes, more unpredictable ways. Actions containing the highest predictive information for an individual's next behavior are negative, such as attacks or enemy markings, while the positive actions of friendship marking, trade and communication contain the least amount of predictive information. However, we show that the system is, to a large part, Markovian and almost devoid of any higher order correlations when taking into account the sign of the action, showing that positive or negative behavior is not more predictable when a longer history of previous actions is accounted for.

The distributions of entropies and predictability found here is strikingly similar to distributions found for datasets of offline mobility \cite{song2010lph}, economic transactions \cite{krumme2010pis}, online conversations and online location check-ins \cite{wang2012hro}, therefore suggesting a possible universality in the limitations of human behavior and its independence of the concrete medium or context. However, contrary to our result of little \mbox{high-order} correlations in behavior, a recent study has shown that the behavior of browsing web pages is, to a large extent, non-Markovian \cite{chierichetti2012awu}. Non-extensive entropies have been recently developed that might be well applied for non-Markovian settings in complex social systems \cite{hanel2011ccc,hanel2012generalized}.

Our observations also provide additional evidence for the fundamental differences in positive and negative behavior that were previously found on dynamic \cite{thurner2012egc} and structural \cite{szell2010mol,szell2012sdl,szell2010msd} levels. Although previously large-scale evidence has confirmed in online human behavior a number of known or hypothesized behavioral phenomena of offline behavior, it is not immediately clear how asymmetries between positive and negative behavior in our, to some extent, artificial, online world can be translated to the offline world. Future research should aim to analyze positive and negative relationships and behaviors that happen in real-life societies and organizations \cite{labianca2006esl}, especially considering the multi-relational aspect of social organization \cite{szell2010mol,kivela2013mn}. Fine-grained datasets of socio-economic behavior, such as the one presented, offer the further possibility of going beyond observations and measurements, \mbox{to study the} mechanisms and origins of behavior in the view of collective phenomena \cite{tadic2013cme}.

\section{Notes added in proof}

During the redaction of this paper, we were made aware of a relevant study that applied the conditional entropy of signed messages to model growth of entropy in emotionally charged online dialogues \cite{sienkiewicz2013eme}.

%%%%%%%%%%%%%%%%%%%%%%%%%%%%%%%%%%%%%%%%%%

\acknowledgements{Acknowledgments}

Roberta Sinatra is supported by the James S.~McDonnell Foundation. Michael Szell thanks the National Science Foundation, the Singapore-Massachusetts Institute of Technology Alliance for Research and Technology (SMART) program, the Center for Complex Engineering Systems (CCES) at King Abdulaziz City for Science and Technology (KACST) and Massachusetts Institute of Technology (MIT), Audi Volkswagen, Banco Bilbao Vizcaya Argentaria (BBVA), The Coca Cola Company, Ericsson, Expo 2015, Ferrovial and all the members of the MIT Senseable City Lab Consortium for supporting the research. Both authors also thank the Santa Fe Institute for the opportunities offered during the Complex Systems Summer School 2010, where some ideas for this project originated.

%%%%%%%%%%%%%%%%%%%%%%%%%%%%%%%%%%%%%%%%%%

\conflictofinterests{Conflicts of Interest}

Michael Szell is an associate of the company, Bayer \& Szell OG, which is developing and maintaining the online game, Pardus, from which the data was collected.

%=================================================================
% References: Variant A
%=================================================================
% Back Matter (References and Notes)
%----------------------------------------------------------
% Style and layout of the references
%\bibliographystyle{mdpi}
%\makeatletter
%\renewcommand\@biblabel[1]{#1. }
%\makeatother
%
%\begin{thebibliography}{1}
%
%% Reference 1
%\bibitem{ref-journal}
%Lastname, F.; Author, T. The title of the cited article. {\em Journal Abbreviation} {\bf 2008}, {\em 10}, 142-149.
%
%% Reference 2
%\bibitem{ref-book}
%Lastname, F.F.; Author, T. The title of the cited contribution. In {\em The Book Title}; Editor, F., Meditor, A., Eds.; Publishing House: City, Country, 2007; pp. 32-58.
%
%\end{thebibliography}

%=================================================================
% References: Variant B
%=================================================================
% Use the following option to include external BibTeX files:
%\bibliography{entropyreferences}

\bibliographystyle{mdpi}

\end{document}